\documentclass[twocolumn,secnumarabic,amssymb, nobibnotes, aps, pre]{revtex4}
\usepackage{graphicx}
\usepackage{amsmath}
\usepackage{epsfig}
\usepackage{color}

\begin{document}

\title{Propagation of Tension along a Polymer Chain}

\author{Payam Rowghanian}
\email{payam.rowghanian@physics.nyu.edu}
\author{Alexander Y. Grosberg}
\affiliation{Department of Physics, New York University, New York, NY}

\begin{abstract}
We study the propagation of tension caused by an external force along a long polymeric molecule in two different settings, namely along a free polymer in 3d space being pulled from one end, and along a pre-stretched circular polymer, confined in a narrow circular tube. We show that in both cases, the tension propagation is governed by a diffusion equation, and in particular, the tension front propagates as $t^{1/2}$ along the contour of the chain. The results are confirmed numerically, and by molecular dynamics simulations in the case of the 3d polymer. We also compare our results with the previously suggested ones for the translocation setting, and discuss why tension propagation is slower in that case.

\end{abstract} 

\maketitle

\section{Introduction}

The behavior of a long polymeric molecule under tension and external forces has been studied for a long time. Blob considerations used to describe the steady trumpet-like conformation of a polymer pulled from one end or under a flow of solvent \cite{BrochardTrumpet,brochard1995polymer} has become a common way of picturing polymer molecules dragged by external forces in a solvent. Also, propagation of the entropic tension caused by an external force has been studied using blob \cite{brochard1994unwinding,doi:10.1021/ma9600769,sakaue2012dragging} and more elaborate considerations \cite{springerlink:10.1140/epje/i2006-10221-y,nam2007kinetics,PhysRevE.80.040801,PhysRevE.83.021802}. Blob considerations have been also used recently to address a more sophisticated dynamical process in which a polymer molecule is forcefully detached from a solid surface \cite{paturej2012polymer}. 

Among other applications, tension propagation has been the core of some of the theories which describe the translocation of a polymer through a nanopore. Some works \cite{PhysRevE.81.041808,RowghanianGrosberg} have used trumpet models to implement the idea that the external force exerted in the pore creates a tension which propagates along the translocating polymer. Taking a different approach, \citeauthor{Panja20081630} \cite{Panja20081630} have considered the propagation of such tensions by accounting for the memory effects that make the motion of the monomers subdiffusive. The diversity of the obtained results for the translocation problem could be associated with different accounts for the tension propagation. 

In this work, we study the tension propagation in two different settings. As the first setting, we consider the formation of a trumpet shape as one end of an equilibrium polymer starts to be pulled by a constant force at time $t=0$. We demonstrate that the tension propagation is governed by a peculiar diffusion equation in which the diffusion coefficient depends on the tension itself. This agrees with a very recent work \cite{sakaue2012dragging} which was done on the same topic. The second setting is an essentially one dimensional chain, fully stretched and confined along a circular tube. At time $t=0$, a weak force exerted at a fixed point along the tube is switched on, and an extra tension caused by the force starts to propagate along the pre-stretched chain. It turns out that in both settings there exists a tension front whose position along the contour of the chain scales with time as $n(t)\sim t^{1/2}$. We compare this result with translocation, and explain why a slower propagation is expected in translocation. 

\section{Trumpet Formation}
A polymer molecule pulled from one end at $t=0$, undergoes conformational changes as it begins to move. Due to the flexibility of the polymer, the force does not immediately affect all the chain, and the information about the force gradually propagates in the form of entropic tension along the chain. After sufficient time has passed, the polymer reaches a steady state where all the monomers move with the same velocity $v$, and the drag force exerted on each monomer by the solvent is balanced by the gradient of the chain tension. Before reaching this steady state, however, the coil essentially consists of $n(t)$ monomers which move under the influence of the force and $N-n(t)$ monomers which do not know about the force yet.

We begin this part with a brief review of the steady trumpet, and then move on to the main topic which is the evolution of the chain from a random coil at time zero into a trumpet. At every moment of time, the $n(t)$ monomers form what we will refer to as a ``non-steady trumpet''. The tension and velocity profiles along this non-steady trumpet are determined by the force balance and mass conservation equations, which, put together, form a diffusion equation for the tension in which the diffusion coefficient is a function of the tension itself. Numerical treatment of this equation confirms the existence of a tension front which separates the moving and immobile parts of the polymer, and propagates as $n(t)\sim t^{1/2}$ along the chain. We also show that although the moving part is a non-steady trumpet, the same scaling behavior could be obtained by approximating the moving part with a steady trumpet. 

Before we proceed, we would like to highlight one point of caveat. In this part, we will make a great use of essential scales and relations which are typically accurate up to numerical factors. Therefore, the equal signs in this part must be really read as ``equal up to numerical factors''. This approach, nonetheless, and like other scaling approaches, reveals much of the underlying physics of the problem. 

\subsection{Steady Trumpet}
\begin{figure}
\includegraphics[width=0.80\linewidth]{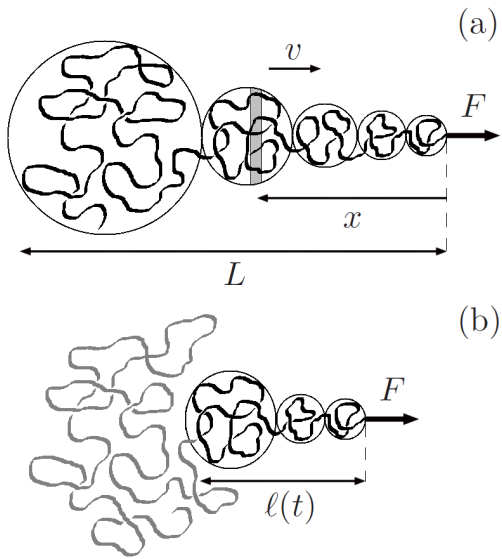}
\caption{a) A steady trumpet of length $L$ moving with velocity $v$. b) A polymer being pulled from one end. $\ell(t)$ is the length along the axis of the non-steady trumpet involved in the motion, beyond which the monomers are unaware of $F$.  \label{trumpet}}
\end{figure}

Let's consider a steady trumpet with $N$ monomers and length $L$ which moves with velocity $v$ (Figure \ref{trumpet}a). This form is created by a moderate force:
\begin{equation}
 N^{-\nu} < \frac{Fa}{T} <1 \label{moderate force}
\end{equation}
which is only strong enough to form blobs which are smaller than the coil size without fully stretching the polymer to its contour length (strong forces create a stem-flower form \cite{brochard1995polymer}). Every monomer in the trumpet feels a drag force $\zeta v$, where $\zeta$ is the friction coefficient per monomer. This drag force has to be balanced by the gradient of the elastic tension $f$, therefore if we consider a slice of width $\delta x$, the force balance will be:
\begin{equation}
 \delta f \sim (\zeta v) \delta N \sim (\zeta v)c(x) \xi^2(x) \delta x
\end{equation}
where $x$ is the position along the axis of the trumpet, $\delta N$ the number of monomers in the slice, $c(x)$ the monomer concentration and $\xi(x)$ the local blob size. 

The friction coefficient is a constant $\zeta_R\sim \eta a$ in the case of Rouse dynamics, where $a$ is the monomer size and $\eta$ the solvent viscosity. In Zimm dynamics, all the solvent inside a blob is dragged with the blob and as a result, the blob feels the friction as a solid whole. Therefore, friction coefficient per monomer depends on the blob size as $\zeta_Z \sim \eta a g^{\nu-1}$, where $g\sim (\xi/a)^{1/\nu}$ is the number of monomers contained in a blob of size $\xi$. It has to be emphasized, however, that strictly speaking, this is only valid for the case of an isolated blob, and is not necessarily locally applicable to a setting in which a group of blobs of different sizes form an extended object. The reason is that each blob can affect the hydrodynamic field around its neighbors, which in turn, changes the friction. In the case of a moving trumpet, however, we find the relation mentioned above to be an acceptable estimation of the friction coefficient. Due to the very nature of the trumpet, the smaller blobs are at the front. The liquid by which the small leading blobs pass is therefore not strongly affected by the larger blobs that will follow. More precisely, a moving blob creates a liquid flow field whose magnitude is only significant up to a distance of order the blob size from the blob. In the case of a trumpet, as we will show, even the size of the largest blob is much smaller than the total length of the trumpet. Therefore, each blob is only considerably affected by the neighboring and not very distant blobs, and thus the friction coefficient can be determined almost locally, as a superposition of the effects of the neighboring blobs. To obtain a mere estimation, we have simply used the local blob size. 

Noting that the blob size depends on the tension as $\xi \sim T/f$ and $c\sim g/\xi^3$, the force balance equation yields:
\begin{subequations}\begin{eqnarray}
 \mathrm{Rouse} \rightarrow \frac{\mathrm{d}f}{\mathrm{d}x}&\sim& -v f^\frac{\nu-1}{\nu} \label{steady Rouse force balance}\\
 \mathrm{Zimm}\rightarrow \frac{\mathrm{d}f}{\mathrm{d}x}&\sim& -v \label{steady Zimm force balance} 
\end{eqnarray}\end{subequations}
where we have made all the quantities dimensionless using the substitutions $f(a/T)\rightarrow f$, $v(\eta a^2/T)\rightarrow v$, and $x/a\rightarrow x$. Letting $f(L)=0$ and $f(0)=F$, the tension profile will be:
\begin{subequations}\begin{eqnarray}
 \mathrm{Rouse}\rightarrow f(x)&\sim&v^\nu (L-x)^\nu \label{steady Rouse force profile}\\
 \mathrm{Zimm}\rightarrow f(x)&\sim& v(L-x) \label{steady Zimm force profile} 
\end{eqnarray}\end{subequations}
We can relate $N$ and $L$ by integrating the concentration, i.e. $N\sim\int_0^L \, c(x)\xi^2(x) \, \mathrm{d}x$. Strictly speaking, the blob size and concentration diverge at $L$, which is meaningless, indicating that the upper bound for such an integration has to be somewhere inside the largest blob. However, the divergence is almost always integrable, except for the case of $\nu=1/2$ under the Zimm dynamics, which can also be regularized as $\left. \frac{\nu}{2\nu-1} (L-x)^{(2 \nu - 1)/\nu} \right|_{\nu=1/2} = \ln (L-x)$. Excluding this case, we have:
\begin{subequations}\begin{eqnarray}
 \mathrm{Rouse}\rightarrow L&\sim&\nu N F^\frac{1-\nu}{\nu} \label{steady Rouse trumpet length}\\
 \mathrm{Zimm}\rightarrow L&\sim&\left(\frac{2\nu-1}{\nu}\right)N F^\frac{1-\nu}{\nu} \label{steady Zimm trumpet length} 
\end{eqnarray}\end{subequations}
where we have used $F=(vL)^\nu$ for the Rouse and $F=vL$ for the Zimm case. Dropping the coefficients we kept above for clarity, the tension profile will be:
\begin{subequations}\begin{eqnarray}
 \mathrm{Rouse}\rightarrow f(x)&\sim&F \left(1-\frac{x}{N} F^\frac{\nu-1}{\nu}\right)^\nu \label{steady Rouse force profile f N}\\
 \mathrm{Zimm}\rightarrow f(x)&\sim&F \left(1-\frac{x}{N} F^\frac{\nu-1}{\nu}\right) \label{steady Zimm force profile f N} 
\end{eqnarray}\end{subequations}

A trumpet is a directed chain whose size along its axis is proportional to its contour length. Moreover, the integration which determines $N$ is not dominated by either the lower or the upper bounds, indicating that the monomers are indeed spread along a length which scales with the contour length. In fact, it is straightforward to determine the number of monomers $N^*$ contained in the largest blob and show that $N^*\ll N$. Noting that the force scale $f^*$ which determines the largest blob size is the same force that balances the drag force exerted on this blob, and after a few lines of algebra, we obtain:
\begin{subequations}\begin{eqnarray}
 \mathrm{Rouse}\rightarrow N^*&\sim&N^\frac{1}{1+\nu} F^{-\frac{1}{1+\nu}} \label{Rouse largest blob}\\
 \mathrm{Zimm} \rightarrow N^*&\sim&N^\frac{1}{2\nu} F^\frac{1-2\nu}{2\nu^2} \label{Zimm largest blob}
\end{eqnarray}\end{subequations}
which is strictly smaller than $N$ for moderate forces (Equation (\ref{moderate force})). 

\subsection{Formation of Trumpet}
Let's now grab one end of an equilibrium coil and start to pull it with a force $F$ at $t=0$. At a later time $t$, a non-steady trumpet of length $\ell(t)$ that contains $n(t)$ monomers forms (Figure \ref{trumpet}b). Since the rest of the polymer does not yet know about the force, the wide end of this trumpet, which is the $n$-th monomer, lies on average in the same place as the original position of the unperturbed coil. Of course as $n$ increases from zero to $N$, this wide end moves a distance comparable to the coil size; however, since we know from the previous part that the extension of the final trumpet scales with $N$, we will neglect this motion which is only on the order of $N^\nu$, and assume that the trumpet end lies at a fixed point in space. 

Unlike the steady trumpet, the velocity is not constant along the non-steady trumpet. It is, however, constrained by a continuity equation, which guarantees that the rate at which the density along the axis changes is determined by the gradient of the flux. Assuming fast blob equilibration, i.e. assuming that the blob scale is determined by the time dependent tension, we can relate the local monomer concentration to the tension and write down the continuity equation in terms of velocity and tension. Together with the force balance relation, we will have a closed set of equations. 

To formulate this, we move to the tip frame, at which the polymer end that is being pulled is at rest. In this frame, the solvent moves with a velocity $u(t)$, and each monomer moves with a velocity $v(x,t)$, where $x$ is the position of the monomer measured from the tip. The force balance equations will then be:
\begin{subequations}\begin{eqnarray}
 \mathrm{Rouse}\rightarrow \frac{\partial f}{\partial x}&=&\left(v(x,t)-u(t)\right)f^\frac{\nu-1}{\nu} \label{formation Rouse force balance}\\
 \mathrm{Zimm} \rightarrow \frac{\partial f}{\partial x}&=&(v(x,t)-u(t)) \label{formation Zimm force balance}
\end{eqnarray}\end{subequations}
which will reduce to Equations (\ref{steady Rouse force balance}) and (\ref{steady Zimm force balance}) at long enough times, when the monomers all have the same velocity as the tip and therefore, $v(x,t)=0$. 

The continuity equation governs the linear monomer density $c\ \xi^2 \sim f^{(\nu-1)/\nu}$:
\begin{equation}
 \frac{\partial}{\partial t}\left(f^\frac{\nu-1}{\nu}(x,t)\right)+\frac{\partial}{\partial x}\left(f^\frac{\nu-1}{\nu}(x,t)v(x,t)\right) =0 \label{continuity}
\end{equation}
Combining the force balance and continuity equations, we obtain:
\begin{subequations}
\begin{eqnarray}
 \mathrm{Rouse}\rightarrow \frac{\partial f}{\partial t}&=&\left(\frac{\nu}{1-\nu}f^\frac{1}{\nu}\right) \frac{\partial^2 f}{\partial x^2}-u(t)\frac{\partial f}{\partial x} \label{diffusion equation Rouse}\\
 \nonumber \mathrm{Zimm}\rightarrow \frac{\partial f}{\partial t}&=&\left(\frac{\nu}{1-\nu}f^\frac{1}{\nu}\right) \frac{\partial^2 f}{\partial x^2}-\left(u(t)+\frac{\partial f}{\partial x}\right)\frac{\partial f}{\partial x}\\
 \label{diffusion equation Zimm}
\end{eqnarray}
\end{subequations}
The velocity and tension at the tip satisfy $v(0,t)=0$ and $f(0,t)={F}$, therefore Equations (\ref{formation Rouse force balance}) and (\ref{formation Zimm force balance}) yield:
\begin{subequations}\begin{eqnarray}
 \mathrm{Rouse}\rightarrow u(t)&=&-\left(F^\frac{1-\nu}{\nu}\right) \left. \frac{\partial f}{\partial x}\right|_{x=0} \label{boundary conditions Rouse}\\
 \mathrm{Zimm} \rightarrow u(t)&=&-\left. \frac{\partial f}{\partial x}\right|_{x=0} \label{boundary conditions Zimm}
\end{eqnarray}\end{subequations}
One last boundary condition, namely $F(L,t)=0$, closes this set of equations. 

What we have obtained is a special kind of diffusion equation in which the diffusion coefficient increases with the tension itself. Therefore, a tension disbalance propagates faster along a chain which is more strongly stretched. The way to understand this is to note that the linear density is lower at places with higher tension, and as a result, the tension has to propagate along a smaller number of monomers to move a distance $\delta x$ along the trumpet axis. The drift term is determined by the solvent velocity, which is nothing other than the tip velocity in the lab frame. $u(t)$ is expected to grow with $F$, therefore as we expect, if we pull harder, the drift term is larger and the tension grows faster. 

Except for a drift term, the equation obtained above is similar to a general form of diffusion equation called Porous Medium Equation (PME), also known as the Nonlinear Heat Equation \cite{vazquez2007porous}. The scaling properties of this equation have been used in a recent work \cite{sakaue2012dragging} on the similar topic as the current work, in which a PME is obtained by considering the problem in the lab frame. This choice of frame on the one hand makes it possible to use the pre-existing knowledge about the PME to characterize the tension propagation, but on the other hand, in this frame, only one boundary condition, namely the tension at the tip, can be explicitly imposed. In the present work, we have chosen the tip frame which results in a more complex equation which includes the tip velocity. The advantage is, however, that the imposition of the two required boundary conditions is straightforward in this frame. This brings in the possibility of obtaining a full analytical or numerical solution to the problem, among which, in this work, we have pursued only the latter.

\begin{figure}
\includegraphics[width=0.99\linewidth]{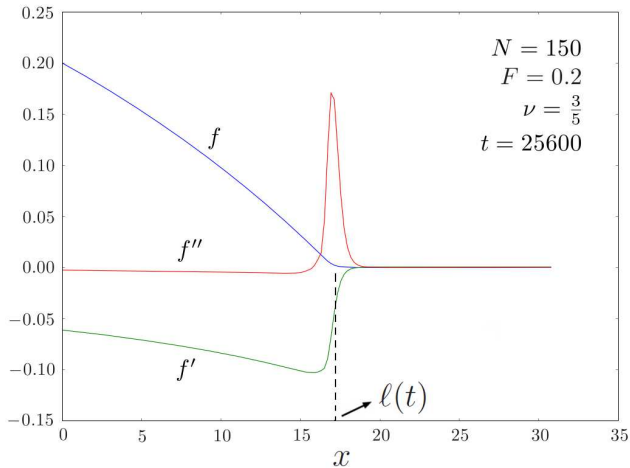}
\caption{Typical form of the tension profile in a non-steady trumpet as a function of position along the trumpet at time $t$. The second derivative of the tension has a large peak at the tension front, $\ell(t)$. Both the first and second derivatives have been magnified for clarity. \label{tension profile}}
\end{figure}

\subsection{Numerical Results}

We treat the diffusion equations obtained above numerically. We set the initial tension profile to be $F$ at the tip and zero elsewhere, and then let the system evolve. What we observe is that at first, the tension propagates along the chain until it reaches the free end of the chain. As expected, this propagation stage is unaffected by $N$, because the tension is unaware of the free end of the polymer. After the tension reaches the free end, it still evolves until it reaches the steady trumpet tension profile.

\begin{figure}
\includegraphics[width=0.80\linewidth]{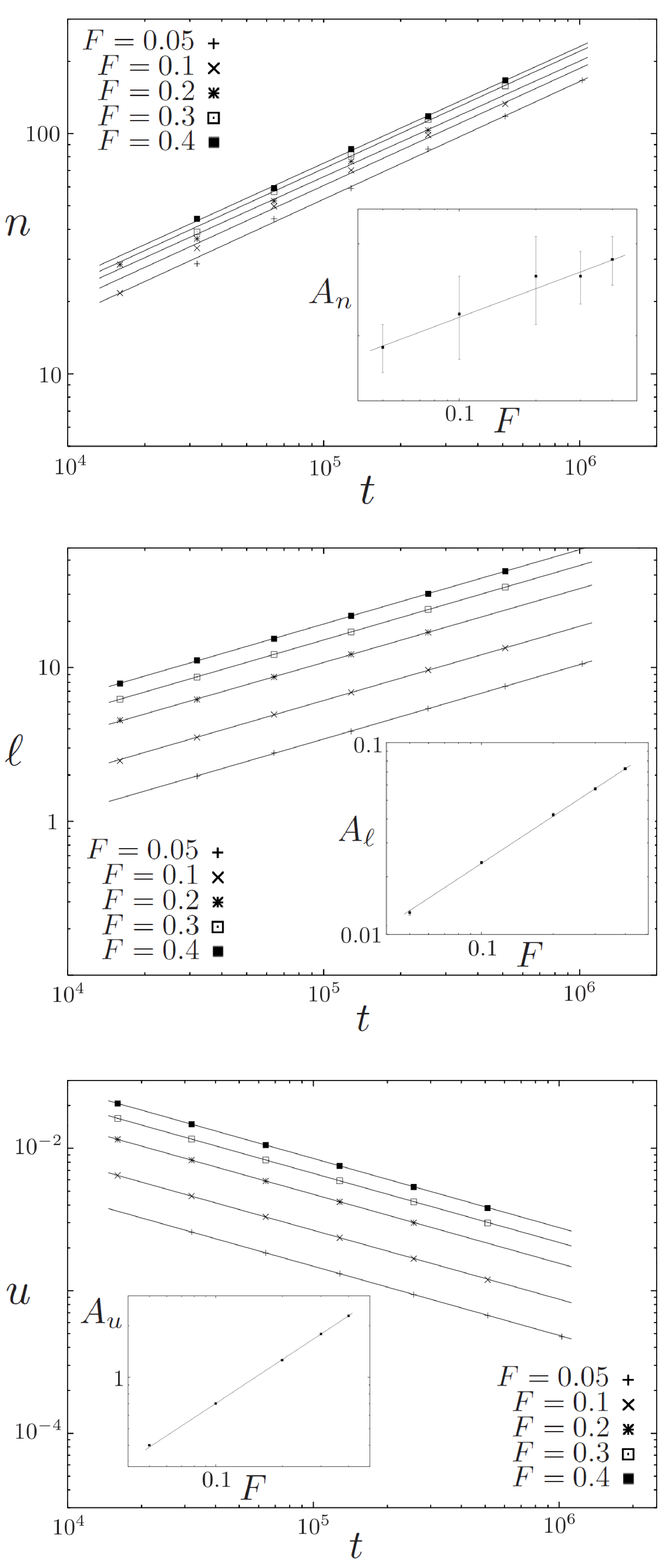}
\caption{Monomer count $n(t)$ and the length $\ell(t)$ of a non-steady trumpet, and the tip velocity $u(t)$, plotted as a function of time for different values of the external force $F$. The force dependence of the same quantities (Equation (\ref{scaling relation})) is shown in the insets. All graphs correspond to the Rouse dynamics for a polymer with $N=150$ and $\nu=3/5$; similar graphs are obtained for other conditions.  \label{exponents}}
\end{figure}

A typical tension profile at some time $t$ is shown in Figure \ref{tension profile}. The second derivative of the tension has a large peak at the point where the tension drops essentially to zero, which supports the idea that the polymer consists of a moving and an immobile part. We use this peak as the tension front and study the scaling behavior of propagation. We are interested in scaling relations of the form:
\begin{equation}
 q(t)\sim A_q(F)\ t^\beta \label{scaling relation}
\end{equation}
where $q(t)$ can be any of the quantities $\ell(t)$, $n(t)$ and $u(t)$, and the amplitudes have the form $A_q(F)\sim f^\alpha$. The graphs in Figure \ref{exponents} demonstrate the form presented above for the case of Rouse dynamics and $\nu=3/5$, and Table \ref{Numerical table} summarizes the values of the exponents $\alpha$ and $\beta$ for $N=150$ at different conditions. The exponents will be compared with the theory developed in the next section.

\begin{table}
 \caption{\label{Numerical table} Numerical Results}
 \begin{ruledtabular}
 \begin{tabular}{|c | c | c | c | } 
 $N=150$ & Rouse $\nu=\frac{1}{2}$ & Rouse $\nu=\frac{3}{5}$ & Zimm $\nu=\frac{3}{5}$ \\ 
 \hline $\alpha_n$ & $0.000$ & $0.182\pm 0.013$ & $-0.24\pm 0.09$ \\ 
 \hline $\alpha_u$ & $0.9986\pm 0.0005$ & $0.85\pm 0.01$ & $0.484\pm 0.003$ \\ 
 \hline $\alpha_{\ell}$ & $0.999\pm 0.005$ & $0.82\pm 0.01$ & $0.49\pm 0.01$ \\ 
 \hline $\beta_n$ & $0.536\pm 0.018$ & $0.487\pm 0.012$ & $0.60\pm 0.03$ \\ 
 \hline $\beta_u$ & $-0.488\pm 0.002$ & $-0.486\pm 0.002$ & $-0.485\pm 0.001$ \\ 
 \hline $\beta_{\ell}$ & $0.489\pm 0.003$ & $0.483\pm 0.002$ & $0.504 \pm 0.003$ \\ 
 %\hline 
 \end{tabular} 
 \end{ruledtabular}
\end{table}

As the graphs show, the largest deviations are observed for $n$. This has to do with the way we calculate $n$, which is to integrate the density from the tip to the tension front. Since the blob size grows very rapidly and to large values near the tension front, $n$ suffers from a relatively larger error. 

\begin{figure}
\includegraphics[width=0.99\linewidth]{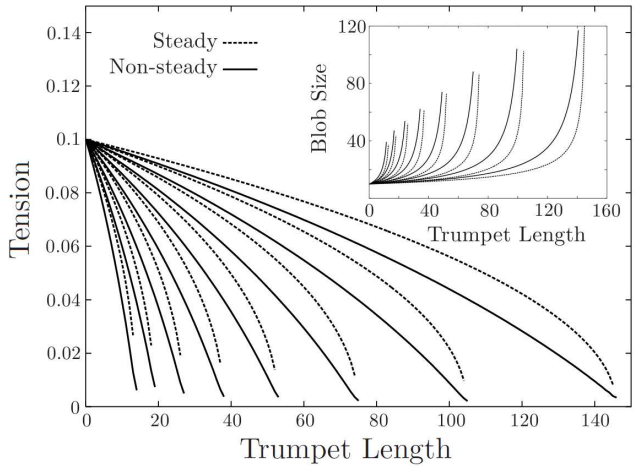}
\caption{Tension profile and blob size as a function of position for non-steady (solid lines) and steady (dashed lines) trumpets. The tension drops somewhat faster in the case of a non-steady trumpet, because it is still in the process of being established along the coil. \label{steady non-steady}}
\end{figure}

The distinction between a steady trumpet and the mobile part of a forming trumpet, to which we have referred as a non-steady trumpet, has been emphasized several times. Figure \ref{steady non-steady} makes an explicit comparison between the two by showing the tension and blob profiles for several pairs of non-steady and steady trumpets of the same lengths. The tension along a non-steady trumpet is consistently smaller than that of a steady trumpet with the same length, showing that the tension has not had sufficient time to grow to its stationary value. Despite the differences, however, the steady and non-steady profiles do not seem to have fundamental qualitative differences. We use this observation in the next section to derive the exponents $\alpha$ and $\beta$. 

\subsection{Derivation of Exponents}
Let's assume that the moving part of the polymer can be approximated by a steady trumpet, characterized by one velocity scale $u(t)$. The rate at which the trumpet length grows is then simply $\dot{\ell}(t)=u(t)$. The friction coefficient of the trumpet is proportional to its monomer counts in Rouse dynamics and to its length in Zimm dynamics. As a result, $F\sim n(t)u(t)$ for Rouse dynamics and $F\sim \ell(t)u(t)$ for Zimm dynamics. Together with Equations (\ref{steady Rouse trumpet length}) and (\ref{steady Zimm trumpet length}) that relate the monomer count and the trumpet length, the following could be obtained:
\begin{subequations}\begin{eqnarray}
 \nonumber \mathrm{Rouse}\rightarrow&\ell(t)&\sim F^\frac{1}{2\nu} t^\frac{1}{2}, \ u(t)\sim F^\frac{1}{2\nu} t^{-\frac{1}{2}},\label{Rouse scaling relations}\\
 &n(t)&\sim F^\frac{2\nu-1}{2\nu} t^\frac{1}{2}\\
 \nonumber \mathrm{Zimm}\rightarrow&\ell(t)&\sim F^\frac{1}{2} t^\frac{1}{2}, \ u(t)\sim F^\frac{1}{2} t^{-\frac{1}{2}},\\
 &n(t)&\sim F^\frac{3\nu-2}{2\nu} t^\frac{1}{2}
\end{eqnarray}\end{subequations}
The relations above are in good agreement with the exponents presented in Table \ref{Numerical table}. The ostensibly indefinite growth of $\ell$ and decay of $u$ with time will be terminated as soon as the tension reaches the far end of the polymer, when both quantities reach their final values. This is of course due to the finite size of the polymer, which is observed in numerical solution (not shown here), but is not captured in the simple scaling analysis developed above.

The only considerable disagreement is observed in the time dependence of $n(t)$ in Zimm dynamics, which has an exponent $0.60$ instead of $1/2$. Therefore, at least for scaling purposes, the non-steady trumpet behaves very similarly to a steady trumpet. Moreover, we can conclude that the tension propagates as $t^{1/2}$ under Rouse dynamics. In the next section, we will find the same relation for a stretched circular coil. 

\subsection{Simulation Results}

\begin{figure}
\includegraphics[width=0.99\linewidth]{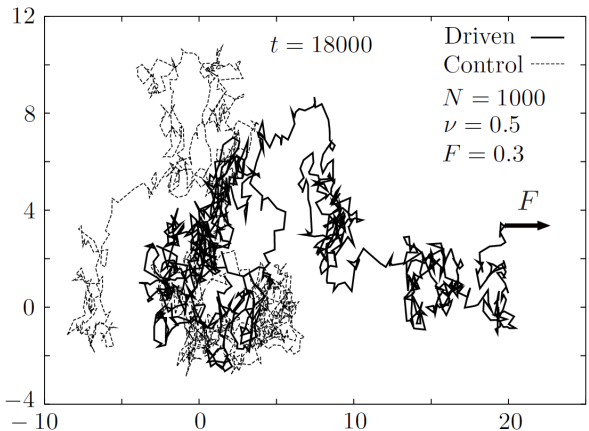}
\caption{Snapshot of a polymer diffusing freely (dashed line) and another being pulled to the left from one end (solid line). A non-steady trumpet has formed in the driven polymer, and the far end which is not yet affected by the force is still in the same position as the non-driven control polymer. \label{snapshot}}
\end{figure}

To test the theory developed above, we have performed some molecular dynamics simulations using a molecular dynamics package called ESPResSo. We have examined the tension propagation along a Gaussian chain of $N=1000$ monomers, which is pulled by a force $F=0.3$ in a non-draining solvent, and is thus governed by Rouse dynamics. The force is measured in units of $T/a$, with $T$ and $a$ both equal to $1$. Since the force dependence of the quantities of interest obtained above is conceivable due to the linearity of the theory, we have focused only on the time dependence in our simulation. Our simulation consists of 550 runs, each of which includes a driven and a non-driven control run which both start from the same random seed. A typical configuration of the system during the simulation is shown in Figure \ref{snapshot}, which can be compared to Figure \ref{trumpet}.

\begin{figure}
\includegraphics[width=0.9\linewidth]{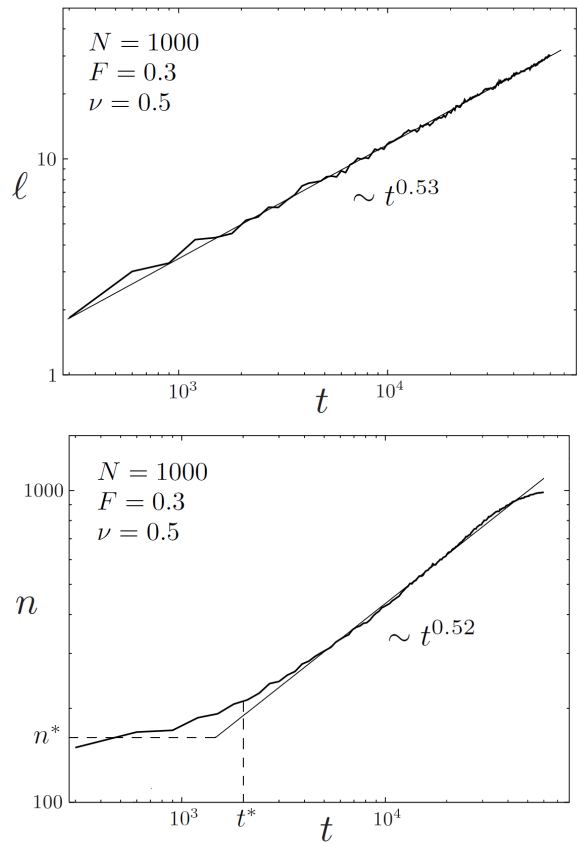}
\caption{Trumpet length $\ell(t)$ and tension front $n(t)$ as a function of time. Both exhibit $t^{1/2}$ scaling, except for $n(t)$ which deviates from the expected behavior at early and final stages of propagation. The deviation turns out to be the result of technical limitations in finding the tension front. For very short times the tension front is hard to find because it is still inside the territory of the equilibrium coil. In the final stages of propagation, the tension front is known to be in the largest blob of the trumpet, and its exact position can not be found using spatial considerations. \label{simulation plots}}
\end{figure}

Trumpet length $\ell(t)$ and tension front $n(t)$ have been plotted in Figure \ref{simulation plots}. While trumpet length exhibits very clearly the expected $\ell(t)\sim t^{1/2}$ behavior, the tension front seems to deviate from the expected $n(t)\sim t^{1/2}$ form for small and large values of $n$. We identify both deviations as consequences of technical limitations involved in finding the tension front in those limits, a point that becomes clear as we describe our method below. Yet another hint on the purely technical nature of the deviations is the close relation between $n$ and $\ell$, among which, the latter is behaving just as expected.

We find the propagation front using the following procedure. At each time, we calculate $R_{cm}$, the position of the center of mass and $R_g$, the radius of gyration of the control polymer. Then, counting from the tip, we find the first monomers whose positions along the direction in which the force is applied are smaller than $R_{cm}-R_g$, $R_{cm}$, and $R_{cm}+R_g$. Finally, we find the average of the indices of these three monomers, and consider it to be the tension front. 

The method described above is expected to be applicable for times when the tip has departed at least a radius of gyration from the polymer's center of mass. Before that, the method simply finds the index of a monomer whose distance from the tip is on average equal to the distance of the tip from the center of mass in equilibrium state, which is the same as the radius of gyration. More precisely, before time $t^*$, this method finds $n^*$ in such a way that $n^*=R_g^2$, which results in $n^*=N/6\approx 160$. Using Equation (\ref{Rouse scaling relations}) and letting $\ell(t^*)=R_g$ we also obtain $t^*\approx 2000$. This is consistent with what we observe in Figure \ref{simulation plots}. 

Before the tension front reaches the far end of the polymer, it propagates along one large blob, which eventually becomes a part of the steady trumpet without any noticeable change in size. Therefore, at the late stages of propagation process, the accuracy of finding the tension front is limited to the scale of the largest blob in the trumpet. Our method, as a result, gives a slower growth at large times, as it effectively calculates the average index of the monomers in the largest blob. 

\section{Tension Propagation in a Circular Chain}

\begin{figure}
\includegraphics[width=0.95\linewidth]{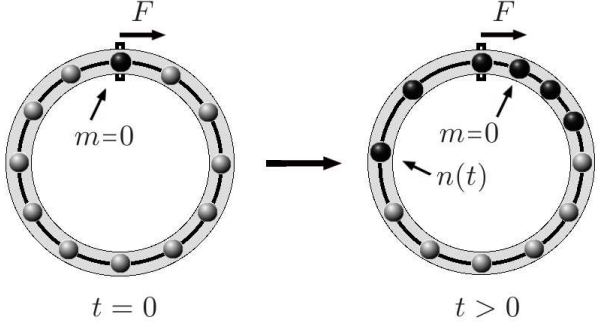}
\caption{A ring polymer confined along a circular tube. At $t=0$, a force is exerted on the monomer labeled $m=0$. By a later time $t>0$, that monomer has passed through the origin, and a few other monomers (shown in black) have also been affected by the force. Due to this motion, the chain at the entrance side is stretched even more strongly than before, while at the exit side, the initial tension has partly relaxed. Therefore, the monomer density per unit length of the tube has decreased at the entrance and increased at the exit side.  \label{circle}}
\end{figure}

Consider a ring polymer, stretched fully along a circular tube of length $L$ (Figure \ref{circle}). This polymer can be considered as a one dimensional system of $N=L/a$ beads and springs, in which the position $x(m,t)$ of a monomer labeled $m$ is measured along the length of the tube. A force $F$, exerted at a fixed point along the tube, to which we will refer as the origin for brevity, is switched on at time $t=0$, and the resulting tension propagates along the chain. Similar to the previous part, and as we will show, the tension has a front which propagates independently of $N$ along the chain as $n(t)\sim t^{1/2}$ before it travels the full length of the chain, i.e. for times when $n(t)\ll N$. 

Neglecting the hydrodynamic interactions, the system can be described by the Rouse dynamics:
\begin{equation}
 \zeta \dot{x}(m,t)=\frac{3T}{a^2}\frac{\partial^2 x}{\partial m^2}+f_r(m,t)-F \delta(s)
\end{equation}
where $\zeta$ is the friction coefficient per monomer, $3T/a^2$ the springs' stiffness, $f_r$ the random Brownian force exerted on each monomer, and $s$ the index of monomers counted from the monomer which is at the origin at time $t$. We are interested in the ensemble average of the equation above, therefore the explicit random Brownian term disappears. However, the randomness is also hidden in $s$, which determines where the external force is exerted. As a result, upon taking the average, the delta-function smears out, which means that the force exertion point itself exhibits a subdiffusive behavior coupled to the dynamics of the rest of the chain. Despite this, and as we will see, the number of monomers driven through the origin scales as $\mathcal{M}(t)\sim t^{1/2}$, while the subdiffusive motion of monomers in a chain (and thus the number of monomers that pass through the origin in this case) scales as $t^{1/4}$. Therefore, the driven motion of the monomers soon becomes the dominant part of the motion and as a result, we neglect the smearing and assume the force exertion point is sharply peaked. Then we obtain:
\begin{equation}
 \zeta \dot{x}(m,t)=\frac{3T}{a^2}\frac{\partial^2 x}{\partial m^2}-F \delta(s)
\end{equation}
where $s=m-\int_0^t v(t') \mathrm{d}t'$ with $v(t)$ being the average monomer passage rate through the origin at time $t$. It is convenient to rewrite this equation in terms of $s$, so that all the monomers get relabeled as each monomer passes through the origin, and the monomer at the origin is always labeled $0$. This yields:
\begin{equation}
 -\zeta v(t)\frac{\partial x}{\partial s}+\frac{\partial x}{\partial t}=\frac{3T}{a^2}\frac{\partial^2 x}{\partial s^2}-F \delta(s)
\end{equation}
which can be written in the following dimensionless form:
\begin{equation}
 -v(t)\frac{\partial x}{\partial s}+\frac{\partial x}{\partial t}=\frac{\partial^2 x}{\partial s^2}-F \delta(s) \label{dimensionless general equation}
\end{equation}
where we have made the substitutions $x/a \rightarrow x$, $F (a/3T)\rightarrow F$, and $v (\zeta a^2/3T)\rightarrow v$. 

After a very long time, the chain moves steadily with a constant monomer passage rate $v$, and a monomer position profile which depends only on $s$. The chain is expected to be more stretched on the entrance side of the origin, and more compact on the exit side. Letting the time dependent term in Equation (\ref{dimensionless general equation}) be zero, we obtain:
\begin{equation}
 -v\frac{\mathrm{d} x}{\mathrm{d} s}=\frac{\mathrm{d}^2 x}{\mathrm{d} s^2}
\end{equation}
where the derivative is discontinuous at the origin:
\begin{equation}
 \left. \frac{\partial x}{\partial s}\right|_{s=0^+}-\left. \frac{\partial x}{\partial s}\right|_{s=N^-}=F
\end{equation}
Adding two other boundary conditions, $x(0,\infty)=0$ and $x(N,\infty)=L$, we find:
\begin{equation}
 x(s,\infty)=N\frac{1-e^{-F\frac{s}{N}}}{1-e^{-F}}
\end{equation}
and $F=Nv$. As expected, $x$ grows faster at smaller values of $s$. 

At $t=0$, the monomers are on average uniformly distributed along the tube and therefore $x(s,0)=s$. When the external force is applied, tension starts to propagate from the origin in both directions. At early stages, meaning before the two tension fronts coincide in the middle of the chain, the tension does not know anything about the size of the chain. Therefore, the early tension propagation is expected to be independent of the chain size and identical for both ring and linear chains. To characterize the tension propagation, we look at different snapshots of the system. Although each monomer may travel several turns along the tube before it reaches the steady state, if we compare snapshots of the system taken at $t=0$ and $t\to \infty$, we find that the position of a monomer which is labeled $s$ changes an amount equal to $y(s,t\to\infty)=x(s,t\to\infty)-x(s,0)$. More generally, the quantity
\begin{equation}
 y(s,t)=x(s,t)-x(s,0)
\end{equation}
can be used to characterize the difference between a snapshot taken at time $t$ and the initial state of the system. Using this quantity, we can now rewrite Equation (\ref{dimensionless general equation}) to obtain:
\begin{equation}
 -v(t)\frac{\partial y}{\partial s}-v(t)+\frac{\partial y}{\partial t}=\frac{\partial^2 y}{\partial s^2}-F \delta(s) \label{dimensionless equation with y}
\end{equation}
The first term is a nonlinear term of order $F^2$, because $v(t)$ and $(\partial y/\partial s)$ are both zero in the absence of the force and therefore must scale with $F$. For small $F$, the nonlinear term is negligible and the equation above will be integrable. This limit corresponds to the same moderate force ($F\ll T/a$ in regular units) considered in the previous part, which is a force too weak to fully stretch the chain. Therefore, given that the chain is fully stretched along the tube in this case, it already feels a tension on the order of $T/a$, and the deformation caused by the external force is small. 

Neglecting the nonlinear term, we obtain a diffusion equation:
\begin{equation}
 \frac{\partial y}{\partial t}=\frac{\partial^2 y}{\partial s^2}+v(t) \label{dimensionless linearized equation with y}
\end{equation}
which has to satisfy the following boundary conditions:
\begin{eqnarray}
 y(0,t)=y(N,0)&=&0 \\
 \left. \frac{\partial y}{\partial s}\right|_{s=0^+}-\left. \frac{\partial y}{\partial s}\right|_{s=N^-}&=&F \label{derivative b c}
\end{eqnarray}
The equation can be solved by taking it to the Fourier space. $y(s,t)$ can only consist of odd modes, as a result:
\begin{equation}
 y(s,t)=\sum_0^\infty y_k(t)\sin\left(\frac{(2k+1)\pi}{N}s\right)
\end{equation}
which can be substituted into Equation (\ref{dimensionless linearized equation with y}) to yield:
\begin{equation}
 y_k(t)=\int_0^t e^{-(t-t')\left(\frac{(2k+1)\pi}{N}\right)^2}U_k v(t') \mathrm{d}t' 
\end{equation}
where $U_k=\frac{2}{(2k+1)\pi}$ is the Fourier component of a constant function whose value is equal to unity. Using Equation (\ref{derivative b c}) we obtain:
\begin{eqnarray}
 \int_0^t&g(t-t')&v(t')\mathrm{d}t'=F \label{memory function b c}\\
 &g(t-t')&=\sum_{k=0}^\infty \frac{4}{N}e^{-(t-t')\left(\frac{(2k+1)\pi}{N}\right)^2}
\end{eqnarray}
where $g(\tau)$ plays the role of a memory function. 

The sum which determines $g(\tau)$ can be calculated for $N\to\infty$. For $\tau\sim N^2$ or larger, the summands decay very rapidly and the sum is therefore dominated by the first term:
\begin{equation}
 g(\tau)\approx \frac{4}{N} e^{-\tau\left(\frac{\pi}{N}\right)^2},\ \ \ \ \ \tau>N^2
\end{equation} 
For smaller values of $\tau$, the sum could be replaced with an integral. It is straightforward to show that this is allowed for $\tau\sim N^\gamma$ with $\gamma<2$, where the difference between two successive summands is very small compared to those summands, and this only changes far through the sum where the summands are already very small. As we will see, we are mostly interested in the time range mentioned above, for which we obtain:
\begin{eqnarray}
 g(\tau)&=&\int_0^\infty\frac{4}{N}e^{-\tau\left(\frac{(2k+1)\pi}{N}\right)^2} \mathrm{d}k\\
 \nonumber &\approx& \frac{1}{\sqrt{\pi \tau}},\ \ \ \ \ \tau<N^2
\end{eqnarray}
We are looking for solutions of the form $v(t)=u\ t^\delta$ for the passage rate, which substituted into Equation (\ref{memory function b c}) will require:
\begin{equation}
  \nonumber \frac{u}{\sqrt{\pi}}t^{\frac{1}{2}+\delta}\int_0^1 \frac{z^\delta}{\sqrt{1-z}}\mathrm{d}z=F
\end{equation}
where $z=t'/t$. Since the left side has to be time independent, we conclude that $\delta=-1/2$. A few lines of algebra will then yield:
\begin{equation}
 v(t)=\frac{F}{\sqrt{\pi t}}
\end{equation}

Putting this all together, we get:
\begin{eqnarray}
 \nonumber y(s,t)&=&\frac{4NF}{\pi^{5/2}}\sum_{k=0}^\infty\left(\frac{1}{2k+1}\right)^2\sin\left(\frac{(2k+1)\pi}{N}s\right)\times\\
 \nonumber && \mathrm{Dawson}\left(\frac{(2k+1)\pi}{N}\sqrt{t}\right)
\end{eqnarray}
where $\mathrm{Dawson}(x)=e^{-x^2}\int_0^x e^{-t^2}\mathrm{d}t$ is the Dawson function. 

\begin{figure}
\includegraphics[width=0.99\linewidth]{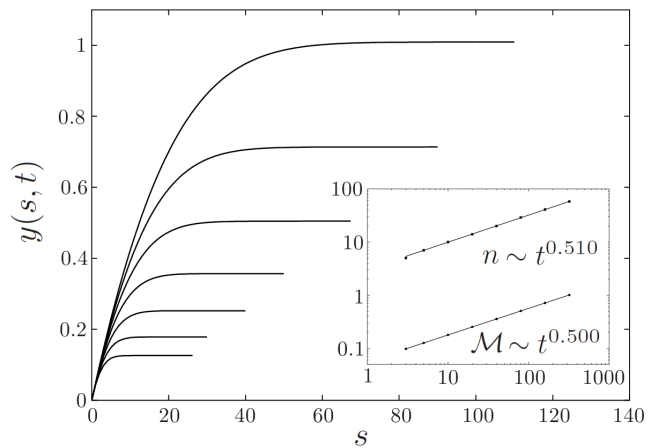}
\caption{Plot of $y(s,t)=x(s,t)-x(s,0)$ at different times, with $x(s,t)$ being the position of a monomer labeled $s$ along the tube at time $t$. Each curve represents one specific time, with the higher curves corresponding to later times. Each plateau is followed by a decay to zero (not shown), which is the reflection of the increasing part shown above around $s=N/2$. The edge of the plateau represents the tension front $n(t)$, and the value of the plateau is equal to $\mathcal{M}(t)$, the number of monomers that have passed through the origin. The inset demonstrates the $t^{1/2}$ scaling for both $\mathcal{M}(t)$ and $n(t)$. \label{y of s}}
\end{figure}

Figure \ref{y of s} shows $y(s,t)$ at different times, which is symmetric around $s=N/2$ (full plot is not shown in the picture). The flat region in the middle corresponds to that part of the chain to which the tension has not propagated yet and therefore, its edge can be considered as the tension front, $n(t)$. The non-zero value of $y(s,t)$ at the plateau is simply due to relabeling all the monomers upon the passage of each monomer through the origin. Therefore, the value of $y(s,t)$ at the plateau must be equal to $\mathcal{M}(t)\sim\int_0^t v(t')\mathrm{d}t'$, which is, as introduced earlier, the number of monomers that have passed through the origin that scales as $\mathcal{M}(t)\sim t^{1/2}$. Figure \ref{y of s} confirms the $t^{1/2}$ dependence of both the tension front and the plateau.

\section{Discussion}

We have shown that the tension caused by an external force propagates as $n(t)\sim t^{1/2}$ in two different settings. This can be attributed to the fact that in both cases, the evolution of the tension is described by a diffusion equation. This result can be compared to the tension propagation in the case of translocation \cite{RowghanianGrosberg}, where we should expect a slower propagation. The reason is that in the presence of the membrane, the driven motion of the monomers is significantly suppressed by the membrane. In particular, the tip of the trumpet moves a distance $\sim N$ before the tension front reaches the end of the polymer, while in the case of translocation, the end that first enters the pore travels only a distance $\sim N^\nu$ before the tension front arrives to the other end. Since the tension propagation is due to the relative motion of the neighboring monomers, the slower departure of the monomers from each other results in a slower tension propagation. The differences mentioned above have been included in our scaling model \cite{RowghanianGrosberg} for the translocation process, in which the important role of the membrane in the dynamics and energetics of the process has been carefully studied. Beyond this, the current models can be improved by including the crowding on the post-translocation side. One way to implement this in the trumpet scenario would be to include the pressure due to crowding in the force balance equations. 

The circular chain and translocation settings share the feature that the force is exerted at a fixed point in space. However, the absence of the membrane and the pre-stretched state of the polymer in the case of a circular chain, causes a dramatic difference between this case and the translocation case. The circular tube confines the monomers to go straight through the origin by suppressing their transverse motion. The moderate force regime considered here corresponds to a condition in which the driving force $F$ is much weaker than the tension created by displacing only one monomer a distance equal to $a$. Since the extra tension created in the chain cannot exceed $F$, for a monomer to be able to pass through the origin, many neighboring monomers have to also move, which is the precise manifestation of the fact that the tension propagates quite far while one monomer passes through the origin. As a result, the fact that the force exertion point moves along the contour of the polymer has a marginal effect, and that explains why the same results are obtained for this case and the trumpet formation case.

While our result is in agreement with the work of \citeauthor{Panja20081630} \cite{Panja20081630}, we believe that both considerations only apply to the cases where monomers can move without any geometric constraints, i.e. when they are far from the membrane. A more elaborate study of the memory function which contains the effect of the membrane is required before the analysis can be applied to the translocation problem.

\end{document}